\documentclass[reprint,aps,prb,amsmath,amssymb,superscriptaddress]{revtex4-1}
\usepackage{dcolumn}   
\usepackage{graphicx}  
\usepackage{bm}        
\usepackage{verbatim}   
\usepackage{amsmath}
\graphicspath{{Figure/}}

\begin{document}
	
	\title{Photoluminescence study of inter-band transitions in few, pseudomorphic and strain-unbalanced Ge/GeSi multiple quantum wells}
	\author{M. Montanari}
	\email[Corresponding author.]{michele.montanari@uniroma3.it }
	\affiliation{
		Dipartimento di Scienze, Universit\`{a} degli Studi Roma Tre, Viale Marconi 446, I-00146 Roma, Italy
	}
	\affiliation{
		IHP - Leibniz-Institut f\"{u}r innovative Mikroelektronik, Im Technologiepark 25, D-15236 Frankfurt (Oder), Germany
	}
	\author{M. Virgilio}
	\affiliation{
		Dipartimento di Fisica “E. Fermi”, Universit\`{a} di Pisa, L.go Pontecorvo 3, I-56127 Pisa, Italy
	}
	\author{C. L. Manganelli}
	\affiliation{
		IHP - Leibniz-Institut f\"{u}r innovative Mikroelektronik, Im Technologiepark 25, D-15236 Frankfurt (Oder), Germany
	}
	\author{P. Zaumseil}
	\affiliation{
	IHP - Leibniz-Institut f\"{u}r innovative Mikroelektronik, Im Technologiepark 25, D-15236 Frankfurt (Oder), Germany
	}
	\author{M. H. Zoellner}
	\affiliation{
		IHP - Leibniz-Institut f\"{u}r innovative Mikroelektronik, Im Technologiepark 25, D-15236 Frankfurt (Oder), Germany
	}
	\author{Y. Hou}
	\affiliation{
		IHP - Leibniz-Institut f\"{u}r innovative Mikroelektronik, Im Technologiepark 25, D-15236 Frankfurt (Oder), Germany
	}   
	\author{M. A. Schubert}
	\affiliation{
		IHP - Leibniz-Institut f\"{u}r innovative Mikroelektronik, Im Technologiepark 25, D-15236 Frankfurt (Oder), Germany
	}   
	\author{L. Persichetti}
	\affiliation{
		Dipartimento di Scienze, Universit\`{a} degli Studi Roma Tre, Viale Marconi 446, I-00146 Roma, Italy
	}
	\author{L. Di Gaspare}
	\affiliation{
		Dipartimento di Scienze, Universit\`{a} degli Studi Roma Tre, Viale Marconi 446, I-00146 Roma, Italy
	} 
	\author{M. De Seta}
	\affiliation{
		Dipartimento di Scienze, Universit\`{a} degli Studi Roma Tre, Viale Marconi 446, I-00146 Roma, Italy
	}
	\author{E. Vitiello}
	\affiliation{
		L-NESS, Universit\`{a} di Milano-Bicocca, Via R. Cozzi 55, 20125 Milano, Italy
	}
	\author{E. Bonera}
	\affiliation{
		L-NESS, Universit\`{a} di Milano-Bicocca, Via R. Cozzi 55, 20125 Milano, Italy
	}
	\author{F. Pezzoli}
	\affiliation{
		L-NESS, Universit\`{a} di Milano-Bicocca, Via R. Cozzi 55, 20125 Milano, Italy
	}
	
	\author{G. Capellini}
	\affiliation{
		Dipartimento di Scienze, Universit\`{a} degli Studi Roma Tre, Viale Marconi 446, I-00146 Roma, Italy
	}
	\affiliation{
		IHP - Leibniz-Institut f\"{u}r innovative Mikroelektronik, Im Technologiepark 25, D-15236 Frankfurt (Oder), Germany
	}
	\date{\today}
	
	\begin{abstract}
		In this work we investigate the structural and optical properties of few, strain-unbalanced multiple Ge/GeSi quantum wells pseudomorphically grown on GeSi reverse-graded substrates. The obtained high epitaxial quality demonstrates that strain symmetrization is not a mandatory requirement for few quantum-well repetitions. Photoluminescence data, supported by a thorough theoretical modeling, allow us to unambiguously disentangle the spectral features of the quantum wells from those originating in the virtual substrate, and to evaluate the impact on the optical properties of key parameters, such as quantum confinement, layer compositions, excess carrier density, and lattice strain. This detailed understanding of the radiative recombination processes is of paramount importance for the development of Ge/GeSi-based optical devices.
	\end{abstract}
	
	
	\maketitle

\section{Introduction}
Ge/GeSi multiple quantum wells (MQWs) have attracted great interest for Si-based photonic devices since the demonstration of quantum confined Stark effect (QCSE) at room temperature (RT)\cite{kuo}. Optical modulators \cite{paul,schaevitz,morini,chaisakulsse,lever}, photodetectors\cite{fidaner,onaran,chang} and spin-based optoelectronic concepts \cite{PhysRevB.85.241303,doi:10.1063/1.4774316,giorgioni,dc} have been investigated using this promising material. Furthermore, RT direct-gap emission has been observed by means of electroluminescence (EL) \cite{chaisakul} and photoluminescence (PL) \cite{gatti}, indicating that Ge/GeSi MQWs are potential candidates for an efficient silicon-compatible light emitter. 
The step further, necessary to exploit Ge/GeSi MQWs as the active medium in a laser, is to obtain positive net optical gain. 

Recently, strain engineering has been proposed to achieve optical gain in Ge/GeSi MQWs \cite{jiang}. Among the several methods investigated to induce external tensile strain in Ge, strategies relying on a silicon nitride external stressor\cite{capellinijap2013,ghrib} are the most promising, since in principle any arbitrary stress can be transfered. Moreover, the combination of external strain and quantum confinement results in two independent parameters to tune the emission wavelength by design, enabling the realization of light emitters of different ``colors" integrated on the same chip. The drawback is that, to minimize the detrimental effect on the emission properties due to an inhomogeneous vertical strain distribution\cite{capelliniopt,chahine}, the thickness of the active region should be limited to a few hundreds of nm. It follows that, in order to expand this strategy to MQWs systems, a limited number of QWs needs to be grown.

To the best of our knowledge, PL of few periods Ge-rich Ge/GeSi MQWs has been demonstrated only on samples grown directly on Ge substrates \cite{chenapl,chennanotechnology}. Unfortunately, heteroepitaxial strain, arising from the 4.2 \% lattice mismatch between Ge and Si, results in crystal defects, such as threading dislocations, that behave as non-radiative recombination centers and their presence is then detrimental for the efficiency of optoelectronic devices \cite{giovane,fiorenza}. 
To achieve high-$x$ Ge$_x$Si$_{1-x}$ layers with low threading dislocations density (TDD), relatively thick GeSi reverse-graded virtual substrates (RG-VS), where the lattice mismatch is gradually distributed among several layers, are commonly used  \cite{capewell,shah,capellinijap107,fitzgeraldapl}.
Employing so many substrate layers with different concentrations results in complicated PL spectra. This is true in particular if a long-wave pump is used for homogeneous excitation and the number of QWs is limited to few repetitions, since both the MQW region and the under-lying layers are simultaneously excited. Indeed in this case a one-to-one identification of all the individual spectral components is not trivial, especially if complementary techniques, such as optical absorption, are not employed.
Thus, for the application of external stressors on few QWs grown on GeSi RG-VS, it is necessary to first systematically characterize the unstressed structures and develop a model to unambiguously assess the optical properties of the Ge MQWs, isolating the emission features due to the excited region of the substrate.\\
Here, we present a study on the optical properties of undoped, unstressed Ge MQWs surrounded by Ge-rich GeSi barriers, grown on relatively thin ($\lesssim$ 2.5 $\mu$m) RG-VS. The high quality of the samples, as probed by high-resolution transmission electron microscopy (HR-TEM), x-ray diffraction (XRD) and atomic force microscopy (AFM), and the good agreement between the observed and simulated data, allowed us to unambiguously interpret the emission spectra.
\section{Sample preparation, experimental and theoretical methods}
The Ge MQWs samples listed in Table \ref{table} have been grown by cold-wall ultrahigh-vacuum chemical vapor deposition (UHV-CVD) on n-Si (001) substrates from ultrapure silane (SiH$_4$) and germane (GeH$_4$), without carrier gas. The pressure during the growth was in the mTorr range while the system base pressure is in the low 10$^{-10}$ Torr range. After an \emph{ex-situ} wet-chemical cleaning, the Si substrates were annealed for 10 minutes in H$_2$ environment at 1150$^\circ$C, to remove the native oxide. To restore a good surface quality, a Si buffer layer has been grown at 850$^\circ$C and $p$= 0.80 mTorr.
Subsequently, we have deposited a 700 nm thick, plastically-relaxed, Ge layer by means of a multi-temperature technique\cite{capellinijap107}.
The rest of the VS has been deposited at 500$^\circ$C, and consists of four Ge$_x$Si$_{1-x}$ layers of 150 nm each, followed by a constant composition buffer layer 1.2 $\mu$m thick. The composition spanned from pure Ge ($x$= 1) to the thick Ge$_{0.81}$Si$_{0.19}$ layer in $\sim$0.05 Ge-composition steps, resulting in a average Ge grading rate of 0.42 $\mu$m$^{-1}$. On top of this VS, we have deposited different Ge wells confined between nominally Ge$_{0.85}$Si$_{0.15}$ barriers with different thickness and/or number of periods. Finally, on top of the MQWs, a 30 nm-thick Ge$_{0.85}$Si$_{0.15}$ cap layer has been deposited. The growth rate of the multiquantum well region was $\sim$ 0.1 nm s$^{-1}$.

The surface morphology of the samples was analyzed by AFM in tapping mode, while TEM was used to study the MQWs structure and the VS.

XRD measurements were carried out with a SmartLab diffractometer from Rigaku equipped with a 9 kW rotating anode Cu source ($\lambda= 0.15406$ nm), a Ge (400)x2 crystal collimator and a Ge (220)x2 crystal analyzer.

Micro-photoluminescence ($\mu$-PL) measurements were carried out using a custom-designed Horiba setup featuring a 50x optical microscope (numerical aperture A= 0.65), a high-resolution spectrometer optimized for IR measurements (Horiba iHR320), an extended-InGaAs detector (0.6 to 1.1 eV detection range), and a liquid nitrogen Linkam cryostat allowing to vary the sample temperature from 80 K up to 350 K with a $\pm$2 K accuracy. 
A 1064 nm laser was focused on the sample surface with a spot size of about 1.7 $\mu m$ and an excitation power density ranging between 5.6$\times$10$^4$ and 5.6$\times$10$^5$ W$\cdot$cm$^{-2}$. All the spectra were collected in backscattering geometry, and a white-body lamp was used to determine the optical response of the set-up used for the spectra calibration.

The electronic band structure and transition energies of the Ge MQWs samples and of the different buffer layers have been calculated relying on two different theoretical framework: a first-neighbor tight-binding Hamiltonian model\cite{Virgilio08,Pizzi10} and a multivalley effective mass description\cite{Wendav}. The predictivity of these two models for the evaluation of electronic spectra in GeSi multilayer heterostructures is well established\cite{virgilioprb,Busby10,Barget17}, and indeed, compatible numerical results have been obtained when calculating the numerical data discussed in this work. 

\section{Results and discussion}
\subsection{Structural analysis}

\begin{table*}
	\caption{\label{table} Material parameters of the investigated samples as determined by TEM and XRD. Samples are labeled as S \emph{thickness of QW-number of periods}.}
	\centering
	\begin{ruledtabular}
		\begin{tabular*}{\textwidth}{@{\extracolsep{\fill}}ccccccc}
			Sample & Ge Thickness & VS Thickness & Periods & $t_w$+$t_b$ TEM (XRD) & $\varepsilon_\parallel$(Ge) & $\varepsilon_\parallel$(GeSi)\\
			Ge & 700 nm & 0 & 0 & / & / & / \\
			VS & 700 nm & 1.8 $\mu$m & 0 & / & / & / \\
			S10-2 & 700 nm & 1.8 $\mu$m & 2 & (26.3) nm & -0.5\% & 0.1 \% \\
			S10-10 & 700 nm & 1.8 $\mu$m & 10 & 10.5+17.2= 27.7 (27.5) nm & -0.6\% & 0.1 \% \\
			S10-5 & 700 nm & 1.8 $\mu$m & 5 & 10.9+17.1= 28.0 (28.1) nm & -0.6\% & 0.1 \% \\
			S17-5 & 700 nm & 1.8 $\mu$m & 5 & 17.1+14.7= 31.8 (31.5) nm & -0.5\% & 0.1 \% \\
			S25-5 & 700 nm & 1.8 $\mu$m & 5 & 25.9+14.6= 40.5 (41.1) nm & -0.6\% & 0.1 \% \\
		\end{tabular*}
	\end{ruledtabular}
\end{table*}

\begin{figure}
	\centering
	\includegraphics[width=\linewidth]{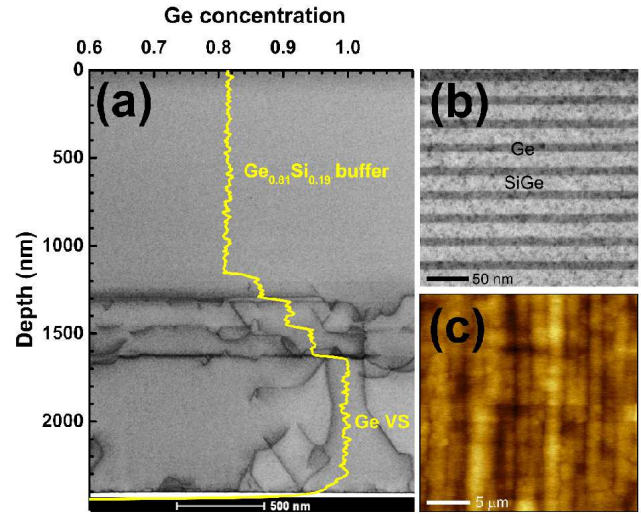}
	\caption{\label{figure1}   (a) EDX composition profile superimposed to the TEM image for the RG-VS of sample S17-5. (b) TEM image of the QWs region of sample S10-10. (c) 25 $\times$ 25 $\mu$m$^2$ surface morphology of sample S10-10 measured with AFM. The height range is 21 nm. Image sides are aligned along the <011>	directions.}
\end{figure}

Figure \ref{figure1}(a) shows the energy-dispersive x-ray spectroscopy (EDX) composition profile superimposed on the corresponding TEM image for sample S17-5, ranging from the interface between the MQWs region and the Ge$_{0.81}$Si$_{0.19}$ buffer down to the Si substrate. No inter-diffusion or segregation is observed within the sensitivity of the technique. \\
In the bottom part of the TEM image, we notice the presence of extended defects, such as threading and misfit dislocations, due to the plastic relaxation of the heteroepitaxial strain \cite{capellinijap107,fitzgerald}.
The threading dislocation density of sample S10-10, obtained by etch-pit counting (not shown here), is about 1$\times$10$^{7}$cm$^{-2}$ on the surface.\\ Homogeneous periodicity and abrupt barrier-well interfaces are observed in the Ge/GeSi QWs stack [Fig. \ref{figure1}(b)]. The thickness of QWs and barriers, evaluated by the analysis of TEM images, is reported in Table \ref{table} as $t_w$ and $t_b$, respectively. A good repeatability of the deposition process is observed, being $t_w$ and $t_b$ nominally the same for samples S10-5, and S10-10.

The lattice tilt, arising from the network of dislocations, leads to a formation of a cross-hatch pattern at the surface \cite{zoellner,acs2015}, as shown in the AFM image reported in Fig \ref{figure1}(c). The root-mean-square surface roughness is about 2.5 nm within a 25 $\times$ 25 $\mu$m$^2$ image.

\begin{figure}
	\centering
	\includegraphics[width=\linewidth]{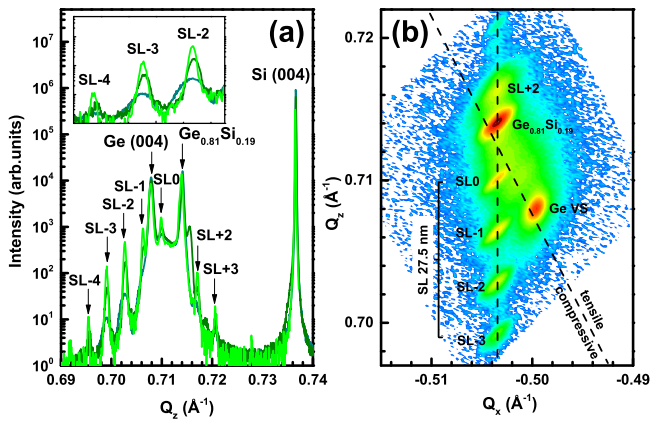}
	\caption{\label{XRD}  (a) XRD rocking curve of samples S10-2 (dark green), S10-5 (green), and S10-10 (light green). The nominal thickness of the QWs is the same. In the inset a detail of SL peaks is reported. (b) Reciprocal space maps of asymmetric ($4\bar{2}\bar{2}$) reflections of sample S10-10.}
\end{figure}

In order to determine the strain status and the actual composition of the GeSi layers, the samples were characterized by XRD rocking curves and XRD reciprocal space maps (RSM).
In Fig. \ref{XRD}(a) we report a (004) rocking curve of samples S10-2, S10-5, and S10-10 around the (004) Ge and (004) Si Bragg peaks. The only difference between the samples is the number of QWs, the thickness of wells ($t_w$) and barriers ($t_b$) being nominally the same. The curve is plotted as a function of the out-of-plane scattering vector $Q_z=\ 4\pi$ sin$(2\Theta/2)/\lambda$. 
Three main peaks are observed at scattering vectors $Q_z$ $\sim$0.708, $\sim$0.714, and $\sim$0.736, which are related to diffraction peaks from the Ge, GeSi and Si layers, respectively. Multiple orders of superlattice (SL) satellites are observed for all the samples, indicating high crystal quality and sharp interfaces between Ge wells and GeSi barriers, as also demonstrated by TEM images. The spacing  between the superlattice fringes (Kiessig fringes) is inversely proportional to the periodicity of the Ge wells \cite{acs2015}, and the spatial periodicity of the grown heterostructures obtained (27.5 nm for sample S10-10) is in good agreement with the analysis of TEM images (27.7 nm for sample S10-10). For all the samples the peak positions are the same, indicating a good repeatability in the thickness of the QWs. As expected, increasing the number of periods the intensity of the SL peaks increases. XRD rocking curve measurements have been carried out on all the samples, and the spatial periodicity obtained is reported in Table \ref{table}. 

HR-XRD reciprocal space maps around asymmetric ($4\bar{2}\bar{2}$) reflections are shown in Fig. \ref{XRD}(b).
The spot corresponding to the Ge$_{0.81}$Si$_{0.19}$ buffer layer is slightly shifted from the relaxation line (i.e., the line of fully relaxed GeSi growth, going from Si to Ge, represented by the dashed diagonal line), indicating that the layer is over-relaxed, due to the difference between the coefficient of thermal expansion in Ge and Si \cite{capellinijap}.
Since the Ge$_{0.81}$Si$_{0.19}$ layer is tensile strained, its in-plane lattice parameter is equivalent to the lattice parameter of a Ge$_{0.86}$Si$_{0.14}$ relaxed bulk alloy.
As a consequence, the MQWs are not strain-compensated.
Nevertheless, Figure \ref{XRD}(b) indicates that the peaks related to the MQWs are vertically aligned to the peak of the Ge$_{0.81}$Si$_{0.19}$ buffer layer (dashed vertical line).
It follows that, although the strain is not symmetrized, the entire MQW stack is coherent with the in-plane lattice parameter of the under-lying VS, thanks to the small number of periods.
Owing to the coherent growth, the Ge wells are tetragonally distorted with an in-plane lattice strain $\varepsilon_\parallel$= -0.6$\pm$0.1 \%, while the Ge$_{0.85}$Si$_{0.15}$ barrier lattice is slightly tensile strained, being $\varepsilon_\parallel$= 0.1$\pm$0.1 \% (see Table \ref{table}).

\subsection{Optical properties}

We now discuss the $\mu$PL properties of the investigated samples. Aiming at an unequivocal identification of the origin of the peaks in the PL spectra, we have first studied the optical properties of two benchmark samples.
The first is the Ge/Si layer and the second is the Ge$_{0.81}$Si$_{0.19}$ VS grown on top of Ge/Si (top two lines in Table \ref{table}), labeled in the following as Ge and VS, respectively.

\begin{figure}
	\centering
	\includegraphics[width=\linewidth]{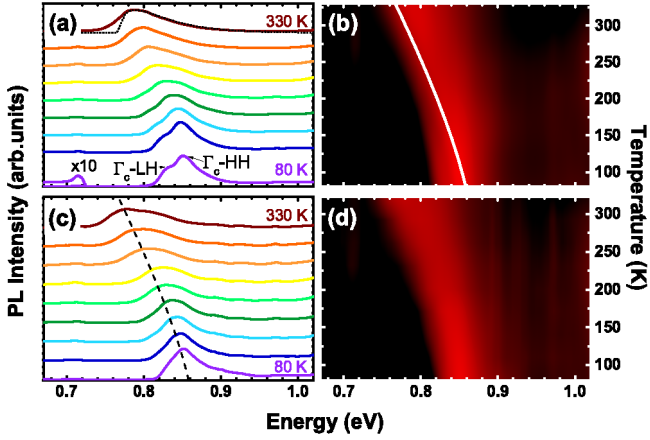}
	\caption{\label{pl1} \emph{Left}: PL spectra acquired on Ge (a), and VS (c). Lattice temperatures range from 80 K to 330 K in $\sim$ 30 K steps. The spectral shape of the direct transition at 330 K, as given in Ref. \onlinecite{bebb}, is also reported as dotted line in the (a) panel. The signal related to the indirect $L_c$-LH transition at 80 K in panel (a) has been enhanced by a factor 10. \emph{Right}: Contour plot of the PL spectra of Ge (b) and VS (d) as a function of $T$ with integrated intensity at each temperature normalized to unity. The results of the fitting of experimental data with Varshni equation are reported as continuous and dashed lines in panels (b) and (c), respectively. }
\end{figure}

In Fig. \ref{pl1},  we report the PL spectra acquired at a constant pump power density of  4$\times$10$^5$ W$\cdot$cm$^{-2}$, and varying the lattice temperature from 80 K to 330 K, in $\sim$ 30 K steps. The spectra acquired on Ge [Fig. \ref{pl1}(a)]  show a broad peak at $\simeq$ 0.85 eV at 80 K, that redshifts to $\simeq$ 0.80 eV at RT. We attribute this feature to the $\Gamma_c$-HH direct band-to-band recombination. PL experimental data associated to $\Gamma_c$-HH transitions have been fitted with the $T$-dependence of the direct-gap, following Varshni equation: 

\begin{equation}
	\label{varshni}
	E(T)= E(0)-\frac{\alpha T^2}{T+\beta}= 0.868-\frac{5.82\times10^{-4} T^2}{T+296},
\end{equation}

where the parameters $\alpha$ and $\beta$ are those of bulk Ge \cite{levinshtein} and $E(T)$ is in eV. The behavior of the peaks as a function of temperature can be clearly observed in the contour plot of the PL spectra, reported in Fig. \ref{pl1}(b), where the fitting of the $\Gamma_c$-HH transitions is shown as a continuous white line. Increasing the temperature, the PL peak broadens and visually redshifts, due to temperature-induced shrinking of the gap.  

For comparison, we also report in Fig. \ref{pl1}(a) the spectral shape of the direct-gap recombination, obtained in the non-degenerate regime, following Ref. \onlinecite{bebb}:

\begin{equation}\label{spectralshape}
	I(\hbar\omega)= \sqrt{(\hbar\omega-E_g)}\cdot\exp(-(\hbar\omega-E_g)/k_BT).
\end{equation}

In Eq. (\ref{spectralshape}), $E_g$= 0.769 eV is the direct gap energy, $k_B$ is the Boltzmann constant, and $T$= 450 K is the temperature of excited carriers, which is found to be higher than the lattice temperature $T_L$= 330 K, due to pump-induced electron heating effects.

On the low-energy side of the peaks acquired at low temperatures, we can see a shoulder related to the $\Gamma_c$-LH direct recombination. As a matter of fact, for moderate in-plane strain the LH-HH splitting $\delta$ is linearly dependent on the biaxial tensile strain $\varepsilon_\parallel$, as $\delta= (6700\pm50)$ meV $\times\ \varepsilon_\parallel$ \cite{carroll,PhysRevB.92.201203}. The in-plane strain arising from the differences in the thermal expansion coefficients $\alpha_i$ between Ge and Si during the cooling process from a high temperature $T_H$ down to a lower temperature $T_L$ is given by:

\begin{equation}\label{thermalstrain}
	\varepsilon_\parallel(T_L,T_H)\simeq \int_{T_L}^{T_H} (\alpha_{Ge}(T')-\alpha_{Si}(T'))\ dT'.
\end{equation}

From Eq. (\ref{thermalstrain}), it is clear that, decreasing the temperature $T_L$, the in-plane  strain $\varepsilon_\parallel$ increases, leading to a larger LH-HH splitting $\delta$. Assuming $T_H\simeq$ 875 K\cite{capellinijap}  and $T_L\simeq\ 80$ K, $\varepsilon_\parallel$ is estimated to be $\simeq$ 0.25 \% at 80 K. The corresponding LH-HH splitting is $\simeq$ 17 meV, compatible with the peak separation in the PL spectra. At 300 K, the biaxial tensile strain $\varepsilon_\parallel$ calculated with Eq. (\ref{thermalstrain}) is reduced to $\varepsilon_\parallel$= 0.17, perfectly matching the value obtained by XRD measurements. The corresponding LH-HH splitting $\delta$ is $\simeq$ 11 meV. Due to the reduced separation as well as the increased electron thermal energy, associated to a larger density of states for the HH band, the peak related to the direct $\Gamma_c$-LH can not be clearly resolved.  Indeed, the relative PL intensity of the two features, in the temperature range investigated, can be explained considering that, although the $\Gamma_c$-LH transition is energetically favored, the final density of states for the $\Gamma_c$-HH recombination and the associated dipole in the out-of-plane direction are larger than the corresponding quantities for the $\Gamma_c$-LH transition \cite{virgilioprb}.

Finally, it is interesting to underline that the signal related to the indirect $L_c-\Gamma_v$ transition is extremely low at all the investigated temperatures [the small signal at $\sim$ 0.71 eV in Fig. \ref{pl1}(a), related to the $L_c$-LH transition, has been enhanced by a factor 10], and PL spectra are dominated by direct recombinations. The rationale is that, since we are dealing with epitaxial thin films, the optical path of the emitted light is small, and then the direct gap emission is not as much reabsorbed as in bulk Ge \cite{arguirov}. Moreover, in the whole investigated temperature range, excess electrons have sufficient thermal energy to populate the $\Gamma_c$ valley, where the recombination rate is much higher \cite{virgiliojap2015}.

The behavior of the PL spectra acquired on the VS sample [Fig. \ref{pl1}(c)] is similar to that observed in the Ge sample but, in this case, the high-energy side of the peaks is broader.
Indeed, the fitting of the $\Gamma_c$-HH transitions in Ge with the Varshni equation (dashed line) reported in Fig. \ref{pl1}(c) clearly evidences the presence of an high-energy shoulder.
Since the direct-gap energy of a SiGe alloy is an increasing function of its Si content, we can attribute this feature to the fact that we are also probing the direct recombination across the VS layers.
In particular, our numerical results indicate that, at the pump-energy used, the 95\% step of the RG-VS also contributes to the PL signal, while the other GeSi layers, richer in Si, remain almost transparent to the excitation (note that, as a consequence, reabsorption effects involving photons emitted from the inner Ge and Ge$_{0.95}$Si$_{0.05}$ layers can be also neglected in our samples).

\begin{figure}
	\centering
	\includegraphics[width=\linewidth]{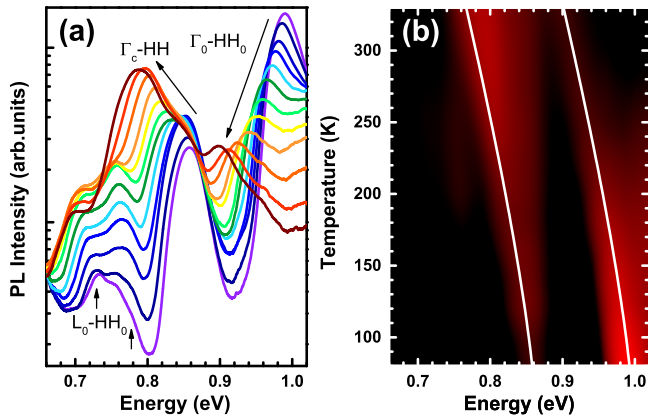}
	\caption{\label{plw} (a) PL spectra acquired on sample S10-10. Lattice temperatures range from 80 K (violet) to 330 K (maroon) in $\sim$ 30 K steps. (b) Contour plot of the PL spectra on sample S10-10 at different lattice temperatures. The integrated intensity at each temperature has been normalized to unity. The results of the fitting of experimental transitions in the Ge layer and the Ge QWs with Varshni equation are also reported as continuous lines.}
\end{figure}

\begin{table*}
	\caption{\label{table2} Experimental and calculated $\Gamma_0$-HH$_0$ and $L_0$-HH$_0$ transition energies at selected temperatures.}
	\centering
	\begin{ruledtabular}
		\begin{tabular*}{\textwidth}{@{\extracolsep{\fill}}cccccc}
			Temperature (K) & \multicolumn{2}{ c }{$\Gamma_0$-HH$_0$ (eV)}  & \multicolumn{3}{ c }{$L_0$-HH$_0$ (eV)}  \\
			\cline{2-3} \cline{4-6}
			& Experimental & Theory & Experimental (ph.em.) & Experimental (ph.abs.) & Theory\\
			80 &	0.990 $\pm$ 0.002	& 0.995 &	0.733 $\pm$ 0.002 &	0.779 $\pm$ 0.002	& 0.753\\
			140 &	0.977 $\pm$ 0.002 &	0.980&	0.716 $\pm$ 0.002 &	0.765 $\pm$ 0.002 &	0.738\\
			200 &	0.958 $\pm$ 0.002 &	0.959 &	0.708 $\pm$ 0.002&	0.758 $\pm$ 0.002 &	0.719\\
			300 &	0.913 $\pm$ 0.002 &	0.916&		& 0.706 $\pm$ 0.002 &	0.682\\
			
		\end{tabular*}
	\end{ruledtabular}
\end{table*}

Once that the origin of the peaks in the PL spectra has been established for the benchmark samples, temperature-dependent PL measurements have been carried out under the same conditions on sample S10-10. Since the photon energy is larger than the direct gap of Ge, but smaller than the direct gap of the Ge$_{0.85}$Si$_{0.15}$ barriers, the \emph{quasi-resonant} excitation of carriers involves holes and electronic states confined in the Ge QWs. 

PL spectra are shown in Fig. \ref{plw}. At first glance, a clear difference between sample S10-10 and the two benchmark samples is observed, consisting in the intense feature at high energy ranging from $\simeq$ 1 eV at 80 K to $\simeq$ 0.9 eV at 330 K. Supported by our numerical model, we relate this feature to a direct transition inside the Ge well between the first confined states in the conduction and valence bands ($\Gamma_0$-HH$_0$). As for the direct recombination in Ge, increasing the temperature, the $\Gamma_0$-HH$_0$ transition is redshifted. At the same time, its intensity is quenched. On the low-energy side, a structure related to the indirect transition $L_0$-HH$_0$ is also observed. Since this structure consists of two features separated by $\sim$ 56 meV (arrows in Fig. \ref{plw}), we attribute them to transitions accompanied by emission and absorption of a longitudinal acoustic phonon, being E$_{LA}$= 28 meV \cite{arguirov}. To support this attribution, we note that, increasing the temperature, the relative intensity of the peak related to phonon absorption is enhanced. The experimental and calculated energies of indirect and direct recombination energies are in excellent agreement as shown in Table \ref{table2} for selected temperatures. 

In between the direct and indirect transitions in the QWs, the direct transition ($\Gamma_c$-HH) due to the under-lying VS is also observed.
To better evidence the behavior, as a function of temperature, of the ratio between the intensity of the PL feature related to the direct transition in the Ge well and the direct transition in the VS, we report in Fig. \ref{plw}(b) a contour plot of the spectral intensity where the integrated intensity of each spectrum has been normalized to unity. From Fig. \ref{plw}(b) it is clear that the intensity of the $\Gamma_0$-HH$_0$ peak is quenched at increasing the temperature, while the intensity of the $\Gamma_c$-HH peak is boosted, the ratio between the two features going from $\simeq$ 4.39 at 80 K to $\simeq$ 0.07 at RT. Despite the $T$-dependent non-radiative recombination dynamics in the substrate and in the QW layers is largely undetermined, the observed behavior of the intensity ratio may suggest that the spatial distribution of the excess carrier density becomes more concentrated in the substrate region at increasing temperature.

\begin{figure}
	\centering
	\includegraphics[width=\linewidth]{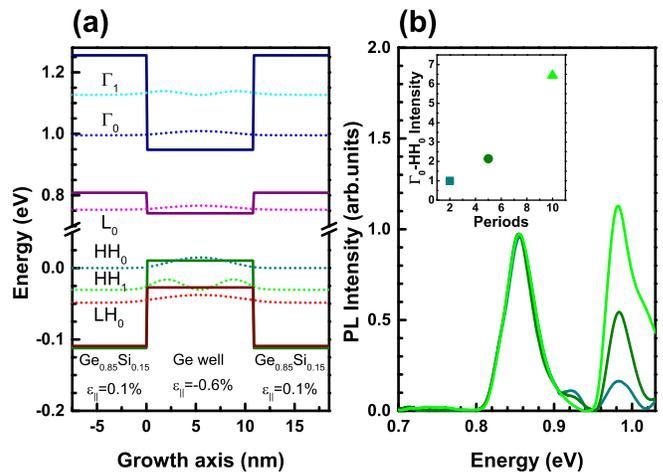}
	\caption{\label{pl2} (a) Conduction- and valence-band edge profiles (continuous lines) and square modulus of the wave functions (dotted lines) for the electron and hole confined states of sample S10-10 at 80 K. (b) PL spectra acquired at 80 K on samples S10-2 (dark green), S10-5 (green), and S10-10 (light green). In the inset: Integrated PL intensity of the QWs peaks as a function of the number of periods.}
\end{figure}

To definitively confirm that the observed high-energy peak is related to the $\Gamma_0$-HH$_0$ transition in the QW, we report in Fig. \ref{pl2}(b) the PL spectra acquired at 80 K on samples S10-2, S10-5, and S10-10 which have equal nominal thickness but different number of periods. Spectra have been normalized so that the intensity of the peak related to the direct transition in the VS is equal to unity. As for the XRD rocking curve, the position of the QW peak is the same for all the samples, indicating good repeatability and the absence of thickness fluctuations, while the intensity of the PL signal is approximately proportional to the number of QW periods, as can be seen in the inset of Fig. \ref{pl2}(b). This observation is compatible with a scenario where the MQWs are uniformly excited and the ratio between the excess carrier density in a single QW and in the substrate does not vary significantly with the number of periods. 

\begin{figure}
	\centering
	\includegraphics[width=\linewidth]{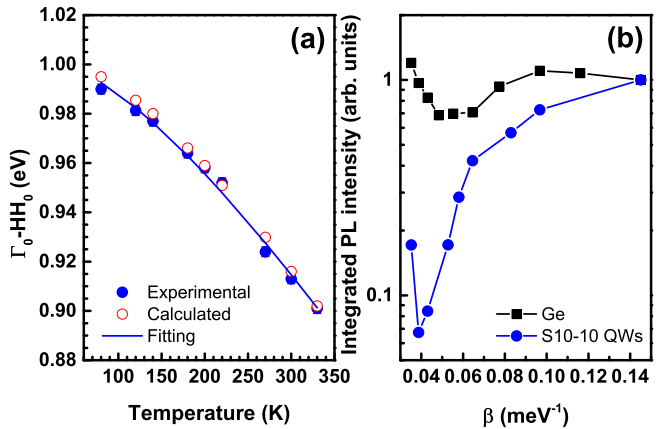}
	\caption{\label{pl3} (a) Experimental (filled circles) and calculated (empty circles) energy of the $\Gamma_0$-HH$_0$ transition in sample S10-10 as a function of the lattice temperature. The fitting of the experimental data with Varshni equation is reported with a continuous line. (b) Integrated PL intensity as a function of $\beta$= 1/$k_BT$. Squares and circles represent Ge and sample S10-10, respectively.}
\end{figure}

To quantitatively characterize the direct band transition in the Ge wells of sample S10-10, we have calculated the corresponding electronic states and band structure [see Fig. \ref{pl2}(a)].
The experimental and theoretical energies for $\Gamma_0$-HH$_0$ as a function of the temperature are reported in Fig. \ref{pl3}(a) as filled and empty circles, respectively.
Experimental data have been fitted following Varshni equation with the same values for $\alpha$ and $\beta$ used in Eq. (\ref{varshni}), but setting a larger $E(0)$ to account for the confinement energy. 
The result of this fitting procedure is reported in Fig. \ref{pl3} as a continuous line.
The experimental, calculated, and fitted energies are in good agreement, confirming that this PL feature originates from direct transitions in the Ge QWs involving the fundamental HH$_0$ and $\Gamma_0$ confined states. 

In the right panel of Fig. \ref{pl3} we show the integrated intensity of the PL spectra as a function of $\beta$= 1/$k_BT$.
The integrated intensity, collected from the Ge sample is displayed as squares, while circles represent the intensity of the QWs feature in sample S10-10.
For both samples, data in Fig. \ref{pl3} have been normalized to unity at 80 K.
The two curves show a non-monotonic trend with a single minimum at $\beta$ equal to 0.048 and 0.039 for the Ge and QW sample, respectively. This behavior can be attributed to the interplay between the thermal boost of the PL intensity induced by the increase of the electron population in the $\Gamma$ valley and by thermal emission of carriers from dislocations \cite{doi:10.1063/1.4955020}, which determines the negative slope in the high-temperature regime \cite{gatti,virgiliojap2015}, and the quenching of the PL dominating in the low-$T$ regime, caused by non-radiative processes whose rate increase with $T$ \cite{virgiliojap, pezzolipra,pezzoliacs}.

\begin{figure}
	\centering
	\includegraphics[width=\linewidth]{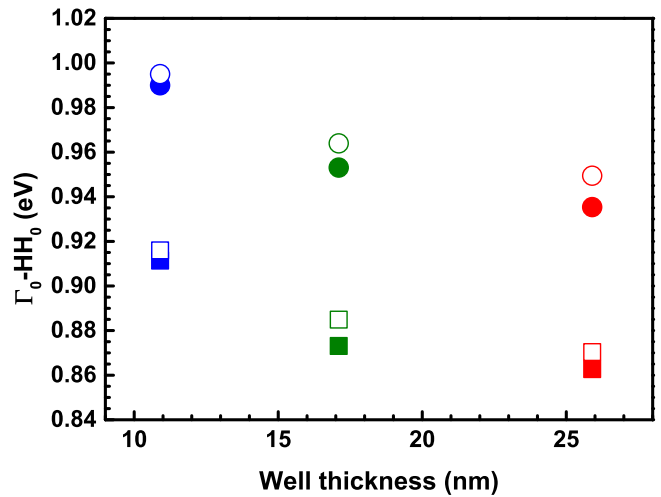}
	\caption{Experimental (filled symbols) and calculated (empty symbols) energy of the $\Gamma_0$-HH$_0$ transition, as a function of the QWs thickness, at 80 K (circles) and 300 K (squares). \label{pl4}}
\end{figure}

To clarify the effect of quantum confinement, we also performed temperature-dependent PL measurements on QWs with different thickness, whose value has been measured by XRD and TEM (see Table \ref{table}).
The experimental and calculated $\Gamma_0$-HH$_0$ transition energies at 80 K and RT are reported as a function of the QW thickness in Fig. \ref{pl4} as filled and empty symbols, respectively.
Their values are larger than the one associated to the direct recombination in the Ge sample ($\simeq$ 0.85 eV at 80 K and $\simeq$ 0.80 eV at RT), due to the concomitant effect of quantum confinement and compressive strain.
Moreover, as expected, a redshift of the PL peak with the increase of the well thickness is clearly observed.\\

\begin{figure}
	\centering
	\includegraphics[width=\linewidth]{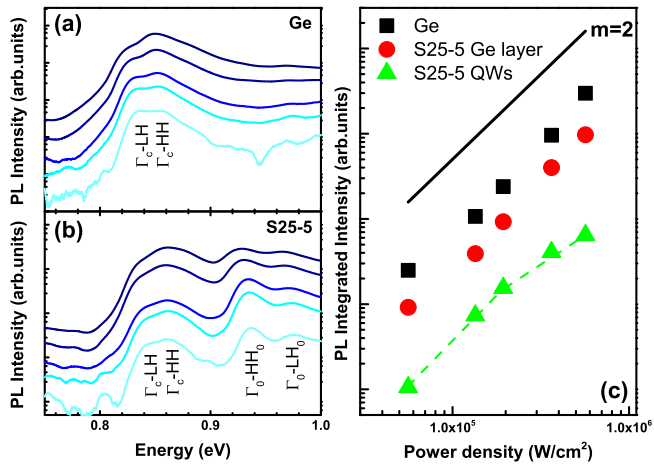}
	\caption{ \emph{Left}: PL spectra measured from the Ge (a) and S25-5 (b) sample at 80 K with different pump power density, ranging in the 5.6$\times$10$^4$-5.6$\times$10$^5$ W$\cdot$cm$^{-2}$ interval. \emph{Right}: Integrated PL intensity for the Ge sample (square), and for the Ge (circle) and QW (triangle) features of the S25-5 sample, as a function of the pump power density. \label{pl5}}
	
\end{figure}

Finally, we conclude discussing PL data collected at different pump-power densities.
In Fig. 8(a), we show PL spectra measured from the Ge sample at 80 K in the 5.6$\times$10$^4$ and 5.6$\times$10$^5$ W$\cdot$cm$^{-2}$ range.
Note that the peak position is not redshifted at high-power density, pointing to the absence of significative pump-induced lattice heating.
The LH-HH splitting is clearly observed in each curve and, increasing the pump power density, the relative intensity of the $\Gamma_c$-HH recombination increases with respect to the $\Gamma_c$-LH one, due to the larger density of hole states.

PL spectra, as a function of the laser pump power density, measured on sample S25-5, with $t_w= 25.9$ nm, are reported in Fig. 8(b).
Again, increasing the power density, the energies of the $\Gamma_c$-HH and $\Gamma_c$-LH recombinations in the Ge layer are not affected, while the peak related to the $\Gamma_0$-HH$_0$ transition slightly redshifts at excitation densities $>$ 1.9 $\times$10$^5$ W$\cdot$cm$^{-2}$.
Interestingly, in this larger-well sample, a spectral feature at $\sim$ 39 meV above the $\Gamma_0$-HH$_0$ one is also distinguishable.
Since our model predicts an excess energy of 40 meV for the $\Gamma_0$-LH$_0$ recombination, we can safely attribute this additional peak to radiative recombinations across the direct gap, involving the light-hole fundamental state. 

Figure 8(c) shows the integrated PL intensity at 80 K as a function of the excitation power density.
Data of the Ge sample and the component related to the Ge layer in S25-5 follow a power-law dependence $I\propto W^m$.
The fit-power exponents $m$ found are close to the theoretical value of $m$= 2 (black line), expected when the dominant non-radiative mechanism is related to SRH recombination \cite{virgiliojap2015}.
On the other hand, the integrated intensity of the QWs feature shows a scaling exponent  $m$= 2 for excitation densities up to 1.9 $\times$10$^5$ W$\cdot$cm$^{-2}$ but, increasing further the excitation, the intensity tends to level off to a value of $m\simeq$ 1.3 indicating the contribution of Auger recombination mechanism.

\section{Conclusions}
In summary, we have analyzed, through micro-photoluminescence measurements and theoretical calculations, the optical properties of undoped, strain-unbalanced Ge MQWs surrounded by Ge-rich GeSi barriers, grown on reverse-graded GeSi virtual substrates by means of ultrahigh-vacuum chemical vapor deposition.
In view of the exploitation of Ge/GeSi MQWs as optical emitters, these results are crucial to unambiguously understand the photoluminescence spectra of samples with few periods of QWs, grown on reverse graded virtual substrates and featuring an external tensile stressor layer.
The high quality of the samples has been confirmed by high-resolution transmission electron microscopy, as well as x-ray diffraction and atomic force microscopy. The structural analysis demonstrates that strain-symmetrization is not a mandatory requirement for few multi-layer repetitions. The good agreement between experimental data and theoretically predicted transition energies, validates the proposed modeling and allows us to distinguish the spectral features originating in the excited portion of the substrate from those associated to the QWs.

\begin{acknowledgments}
	Part of this work is supported by the European Union research and innovation programme Horizon 2020 under grant No. 766719 - FLASH Project.
\end{acknowledgments}

\end{document}